\documentclass[titlepage]{csetr}
\usepackage{alltt}
\usepackage{times,amsmath,epsfig}
\usepackage{lscape}
\usepackage{blkarray}
\usepackage{multirow}
\usepackage{graphicx}
\usepackage{colortbl}
\usepackage{subfigure}
\usepackage{amsmath}
\usepackage{algorithm}
\usepackage{algorithmic}
\usepackage[table]{xcolor}

\definecolor{lightgray}{gray}{0.9}

\newcounter{dc}
\renewcommand{\and}{\hspace{.5cm}}

\trnumber{201220}
\trdate{July 2012}
\title{%
Detecting, Representing and Querying Collusion in Online Rating Systems
}
\author{%
  Mohammad Allahbakhsh$^1$ \and %
  Aleksandar Ignjatovic$^1$ \\ %
  Boualem Benatallah$^1$ \and %
  Seyed-Mehdi-Reza Beheshti$^1$\\ %
  Norman Foo$^1$\and
  Elisa Bertino$^2$\\[2em]
  $^1\, $University of New South Wales\\ Sydney 2052, Australia \\%
  \email{\small{\{mallahbakhsh,ignjat,boualem,sbeheshti,norman\}@cse.unsw.edu.au}}\\ \\
  $^2\,$Purdue University, West Lafayette, Indiana, USA \\%
  \email{\small{bertino@cs.purdue.edu}}\\[3cm]
}

\date{}
\begin{document}
\maketitle

\begin{abstract}
Online rating systems are subject to malicious behaviors mainly by posting unfair rating scores. Users may try to individually or collaboratively promote or demote a product. Collaborating unfair rating 'collusion' is more damaging than individual unfair rating. Although collusion detection in general has been widely studied, identifying collusion groups in online rating systems is less studied and needs more investigation. In this paper, we study impact of collusion in online rating systems and asses their susceptibility to collusion attacks. The proposed model uses a frequent itemset mining algorithm to detect candidate collusion groups. Then, several indicators are used for identifying collusion groups and for estimating how damaging such colluding groups might be. Also, we propose an algorithm for finding possible collusive subgroup inside larger groups which are not identified as collusive. The model has been implemented and we present results of experimental evaluation of our methodology.
\end{abstract}

\section{Introduction}

\subsection{Motivation and Background}

Web 2.0 technologies have connected people all around the world. People can easily contact each other, collaborate on doing jobs or share information through mass collaborative systems which also are called crowdsourcing systems \cite{cswww}. One major group of crowdsourcing systems are used for rating products on the web which are called \emph{Online Rating Systems}. Online rating systems help people judge the quality of products. Since the number of providers and their products is extremely large and ever growing, it is impossible for users to base their choices on their trust in providers they are already familiar with. For that reason the users usually look for the opinions and feedbacks collected from other users who have used or purchased the product before. Providers advertise their products in online rating systems and customers rate them based on their experiences of dealing with that particular product. Based on the rating scores received from customers, the system builds a rating score for every product representing the quality of the product from customers' point of view. Yelp\footnote{http://www.yelp.com}, IMDb\footnote{http://www.imdb.com/} and Amazon\footnote{http://www.amazon.com/} are some of the popular online rating systems.

The big issue in these systems is the trustworthiness of cast feedback. Many pieces of evidence \cite{ebayProblem,amazonproblem} show that users may try to manipulate the ratings of the products by casting unfair rating. \emph{Unfair ratings} are rating scores which are cast regardless of the quality of the product and usually are given based on personal vested interests of the users. For example, providers may try to submit supporting feedback to increase the rating of their product and consequently increase their income \cite{amazonproblem} . The providers also may attack their competitors by giving low scores in their feedback on their competitor's products. Another study shows that sellers in eBay boost their reputations unfairly by buying or selling feedback \cite{ebayProblem}.

Unfair ratings are broadly divided into two categories \cite{dishonest,reliable}~: (i) \emph{individual} and (ii) \emph{collaborative}. Individual unfair ratings are rating scores which people give unfairly without collaborating with others. Such ratings are given for several reasons: lack of expertise, dishonesty or irresponsibility of the user, in the form of, for example, random choice of their rating. Several models have already been proposed for identifying and eliminating these types of scores.

Collaborative unfair ratings are a result of a group of users who try to manipulate the rating of a product collaboratively. This type of unfair rating is usually planned by the product owner or her competitors. The collaborative unfair ratings which are also called \emph{collusion} \cite{IEEE2102Survey,reliable} by their nature are more sophisticated and harder to detect than individual unfair ratings \cite{IEEE2102Survey}. For that reason, in this work we focus on studying and identifying collaborative unfair ratings.

Collusion detection methods are widely studied in P2P systems \cite{CollusionInP2P,eigentrust,simplified}. In P2P systems, users have some pairwise relations which simplify finding collusion between them. In contrast with P2P systems, in online rating systems generally reviewers have no such pairwise relations. Therefore, finding collusive groups in these systems requires other indicators rather than direct relation. Similarity of cast rating scores on same products, time window in which those ratings are cast, deviation from majority, etc are some indicators which are used for detecting collusion groups. \cite{mainWWW,CIKMInds}.

\subsection{Problem Definition}
\label{sec:pd}
While there already are rater sophisticated collusion detection systems, such systems still face some unresolved challenges. The first problem is that most of the existing collusion detection approaches focus on planned collusions, i.e., they suppose that there are some known teams in the systems, for example, the providers of a product,  which may try to collude following some pre-determined collusion plan \cite{thesis, reptrap}. It is obvious that in an online open-to-all crowdsourcing system this model is not applicable.

The second challenge arises when a group of reviewers try to completely take control of a product i.e., when the number of unfair reviewers is significantly higher than the number of honest users;  the existing models usually can not detect such a group. Also, the existing models do not perform well against intelligent attacks, in which group members try to give an appearance of honest users. For example, typically they will not deviate from the majority's ranking on most of the cast feedback and target only a small fraction of the products. Such attacks are hard to identify using the existing methods \cite{reptrap}.

The next issue is that all existing methods check a group as a whole, i.e., they calculate collusion indicators for the group and check its degree of collusiveness. There are cases in which a large group of people have rated the same group of products, and thus could be considered as a potential collusion group. Based on other indicators, such group might be subsequently deemed as non collusive. However, in such cases there may exist some smaller collusive sub-groups inside the large group which are collusive, but when put along with others they might be undetected. Detection of such sub-groups is not addressed in the existing collusion detection models.

\subsection{Contribution and Outline}
Collusion of groups is best represented using collusion graphs, and such graphs are best used for collusion detection by means of a rather generic query language. To the best of our knowledge, there are no easy to use, reusable tools for collusion detection based on graph model representation of rating data collected form online systems and their query using a simple language. In this paper we propose such a model for detecting collusion groups employing six indicators. Four indicators are used for assessing the degree of collusiveness of a group and two indicators to assess how potentially damaging the identified group might be. Our algorithm employs existing clustering techniques, such as Frequent Itemset Mining (FIM) technique \cite{FIM}. The novelty of this work is in a new model, some new collusion indicators and an associated query language. In summary, we present here:

\begin{itemize}
  \item
        A novel approach for analyzing behavior of the reviewers in an online rating system. Besides indicators used in the already existing work, we also define two novel indicators: \begin{enumerate}
        \item an indicator which we call \emph{Suspiciousness} of a reviewer, which is a metric to estimate to what extent ratings posted by such reviewer correspond to majority consensus. Such indicator is calculated using two distance functions: the \emph{$L^p$ distance} and the \emph{Uniform distance}.
        \item an indicator we call \emph{Spamicity} which estimates the likelihood that a particular rating score given to a product by a reviewer is unfair.
        \end{enumerate}
  \item
        We propose a framework for collusion detection in online rating systems. The proposed framework uses an algorithm called RanKit and six indicators for collusion detection. RankIt uses two sub-algorithms which are built on FIM technique for finding potential collusion groups.
         \item
        We propose a new notion and presentation for collusion groups called \emph{biclique}. A biclique is a group of users and a group of products such that every reviewer in such a group has rated every product in the corresponding group of products.
  \item
        We propose a new metric called Damaging Impact (DI) to represent how potentially damaging the identified biclique is. DI is calculated based on the number of reviewers involved in collusion and number of products which have been targeted.

  \item
        We propose a graph data model for representing rating activities in online rating system. This model allows: (i) representing products, reviewers and the rating scores reviewers have cast on products and (ii) identifying bicliques of colluders.

 \item Such identification is effected using a query language which we propose, used for obtaining information on such rating graphs.
\end{itemize}

The remainder of this paper is organized as follows: In section~\ref{sec:prel}  we define and formulate the problem of collusion detection. In section~\ref{sec:cluster} we propose our method of finding candidate collusion groups. Section~\ref{sec:Ind} presents collusion indicators. In section~\ref{sec:sub} we show how we find collusive sub-groups. Section~\ref{sec:frm} presents our proposed collusion detection framework. In section~\ref{sec:eval} we propose implementation details and also evaluate results. We discuss related work in section~\ref{sec:rel} and conclude in section~\ref{sec:Concl}.

\begin{figure*}[!t]
\centering
\subfigure[Sample of Available Products]{
\includegraphics[scale=0.45]{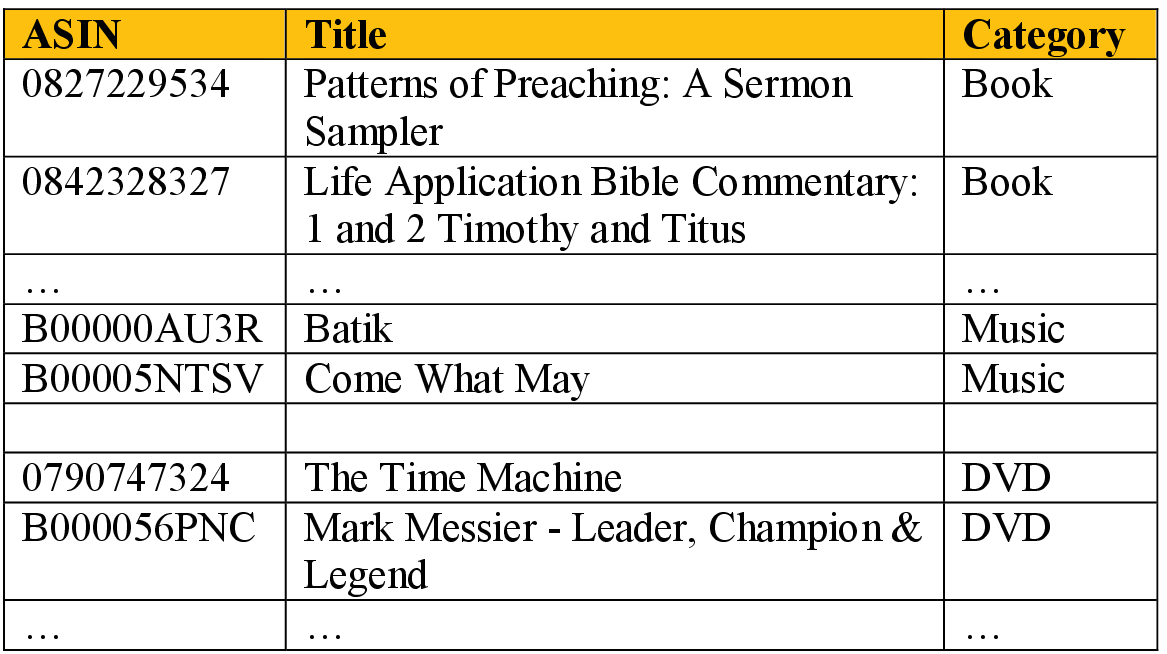}
\label{fig:Products}
}
\subfigure[Some Ratings Cast on Products]{
\includegraphics[scale=0.45]{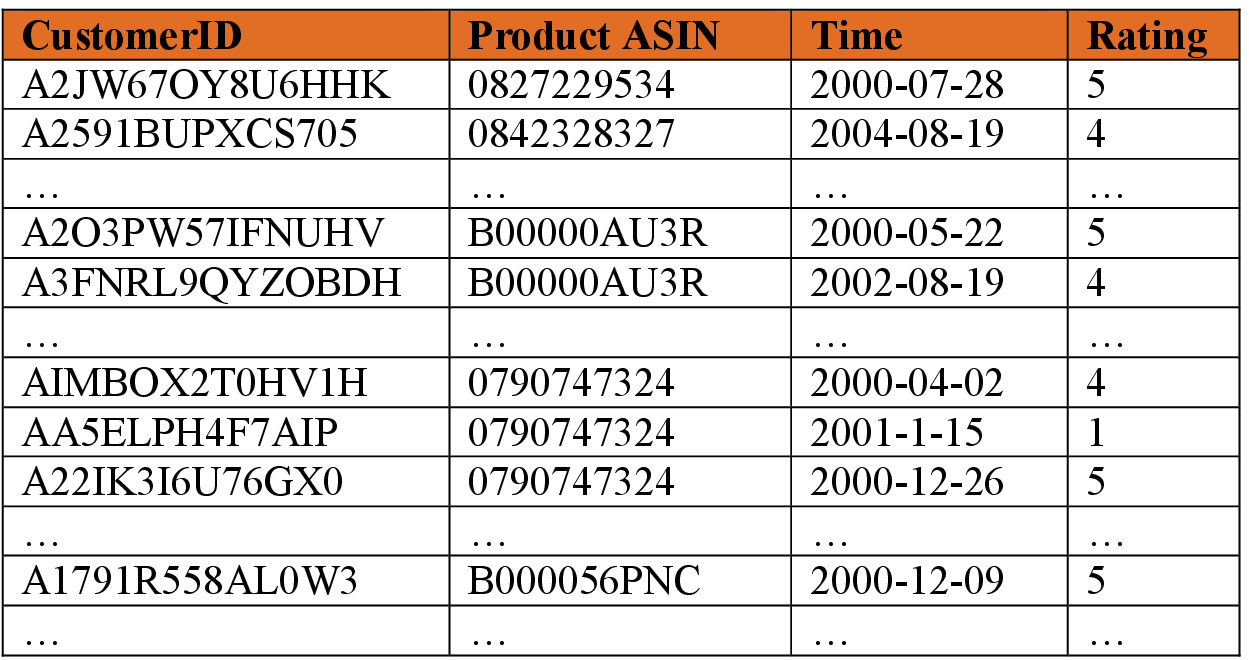}
\label{fig:Rates}
}
\label{fig:Sample}
\caption[]{A Sample of AMZLog}
\end{figure*}

\section{Preliminaries} \label{sec:prel}

In this section we present: (i)~an example scenario for online rating systems; (ii)~a graph data model for representing online rating systems; and (iii)~the process for generating proposed graph model from online rating system logs.

\subsection{Example Scenario}

Amazon is one of well-known online markets. Providers or sellers put products on the Amazon online market. Buyers go to Amazon and buy products if they find them of an acceptable quality, price, etc. Users also can share their experiences of the products with others as reviews or rating scores they cast on products. Rating scores are numbers in range $[1,5]$. Amazon generates an overall rating rank for every product based on the rating scores cast by users. There is evidence (see \cite{amazonproblem}) showing that the Amazon rating system has widely been subject to collusion and unfair ratings.

We use the log of Amazon online rating system\footnote{http://snap.stanford.edu/data/amazon-meta.html} which was collected by Leskovec et. al for analyzing dynamics of viral Marketing\cite{datasetcollector}, referred in the following as AMZLog. This log contains more than $7$ million ratings cast on the quality of more then $500$ thousands of products collected  in the summer of $2006$. Figure \ref{fig:Sample} illustrates a sample of data stored in AMZLog.

\subsection{Data Model}

We define a graph data model (i.e. ORM: Online Rating Model) for organizing a set of entities (e.g. reviewers and products) and relationships among them in an online rating system.
ORM can be used to distinguish between fair and unfair ratings in an online rating system and helps build a more realistic ranking score for every product. In ORM, we assume that interactions between reviewers and products are represented by a directed graph $G = (V,E)$ where $V$ is a set of nodes representing entities and $E$ is a set of directed edges representing relationships between nodes. Each edge is labeled by a triplet of numbers, as explained below.

\subsubsection{ORM Entities}

An entity is an object that exists independently and has a unique identity. ORM consists of three types of entities: products, reviewers and bicliques.

\noindent \textbf{Product.}
A product is an item which has been put on the system to be rated by system users in terms of quality or any other possible aspects. Products are described by a set of attributes such as the unique indicator (i.e. ID), title, and category (e.g. book, cd, track, etc). We assume that there are $N_p$ products $P=\{p_j | 1 \leq j \leq N_p\}$ in the system to be rated.

\noindent \textbf{Reviewer.}
A reviewer is a person who rated at least one product in the system. Reviewers are described by a set of attributes and are identified by their unique identifier (i.e. ID) in the system. We assume that there are $N_u$ reviewers $U=\{u_i | 1 \leq i \leq N_u\}$ rating products in an online rating system.

\subsubsection{ORM Relationships}

A relationship is a directed link between a pair of entities, which is associated with a predicate defined on the attributes of entities that characterizes the relationship. We assume that no reviewer can rate a product more than once. So, in ORM model, there is at most one relation between every product and reviewer. We define only one type of relationship in ORM: \emph{Rating Relationship}.

When a reviewer rates a product, a rating relationship is established between corresponding reviewer and product. We assume that $R$ is the set of all rating relationships between reviewers and products i.e. $e_{ij}$ is the rating the $u_i$ has given to $p_j$. A rating relation is labeled with the values of the following three attributes:
\begin{enumerate}
  \item
        \textbf{\emph{value:}} The value is the evaluation of the the reviewer from the quality of the product and is in the range $[1,M], M > 1$. $M$ is a system dependent constant. for example in Amazon, rating scores reside between $1$ and $5$. We denote the value field of the rating by $e_{ij}.v$.
  \item
        \textbf{\emph{time:}} The time in which the rating has been posted to the system. The time field of the rating is denoted by $e_{ij}.t$.
  \item
        \textbf{\emph{Spamicity:}} As mentioned earlier, we assume that every reviewer can be in a unique relation with each product (i.e., we do not allow multi-graphs).  However, in real world, e.g. in Amazon rating system, one reviewer can rate a product several times. Some collusion detection models like \cite{CIKMInds} just eliminate duplicate rating scores. We rather use them for the purpose of detecting unfair ratings. \emph{Spamicity} shows what fraction of all scores cast for this particular product are cast by the reviewer in the considered relationship with this product. We denote the Spamicity of a rating relation by $e_{ij}.spam$. This will be detailed in section \ref{ssec:pp}.
\end{enumerate}
Clearly, the rating relationship may be labeled with other optional attributes.

\subsection{biclique}

A biclique is a sub-graph of ORM containing a set of reviewers $R$, a set of products $P$ and their corresponding relationships $Rel $. All reviewers in $R$ have rated all products in that $P$, i.e., there is a rating relationship between every $r \in R$ and every $p \in P$. A biclique is denoted by $CL=\{R, P, \}$.

\begin{figure}[!t]
\centering
  \includegraphics[scale=1.5]{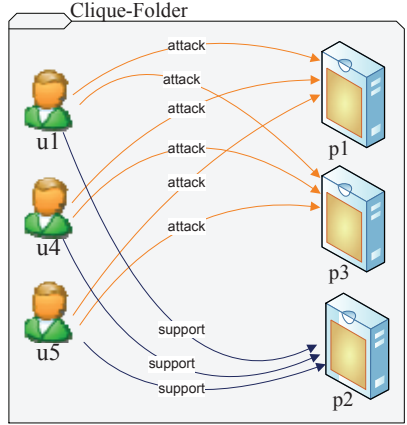}\\
  \caption{An example of a biclique.}\label{folder}
\end{figure}

For example, $CL=\{\{r_1, r_4, r_5\}, \{p_1,p_2, p_3\}, \{r_1 \xrightarrow{} r_1, ... \} \}$ means that reviewers $r_1$, $r_4$, and $r_5$ all have voted on  $p_1$,$p_2$ and $p_3$. Figure~\ref{folder} illustrates this biclique.

\subsection{Data Preprocessing}
\label{ssec:pp}

Since we are proposing a specific graph data model for online rating systems, we have to process data logs and prepare them to fit in our data model. Therefore, we apply several preprocessing steps as data preparation step.

\begin{itemize}
  \item
        We delete inactive reviewers and unpopular products. We suppose that if a reviewer has rated just a few products, she can not be a threat to the reliability of the system, even if she has participated in some collusive activities. For example, we deleted from the data log  \emph{AMZLog} all reviewers who reviewed less than $10$ products. Also, the products with only a few ratings are not attractive for users, so there is a very low chance for a collusion on such a product. So, we delete all such unpopular products; in case of \emph{AMZLog}, we have deleted the products on which less than $10$ ratings are cast.
  \item
        We found in the AMZLog some products which were rated several times by the same user. Posting multiple reviews for a product is common \cite{CIKMInds} and due to several reasons. For example, a buyer may rate a product low because she is not able to use it properly, but after reading manuals or receiving support, she changes her mind and rates the product highly [cikm10].  However, posting a large number of votes on the same product, is a strong indicator of a suspicious behavior of the user. When a user has posted several rating scores we choose the last one as the final opinion of the user about the product. To distinguish between such rating and the normal single ratings, we also record the spamicity value of the rating relationship between the reviewer and product. Thus, let that $E(j)$ be the set of all ratings given to product $j$ and $E(i,j)$ is the set of all ratings given by reviewer $i$ to the product $j$. We calculate spamicity degree of the relationship between the reviewer $r_i$ and the product $p_j$ using the equation \ref{eq:spam}:

        \begin{equation}
        \label{eq:spam}
           e_{ij}.spam = \left\{
          \begin{array}{l l}
            0 & \quad \text{if $(|E(i,j)| \leq 2)$}\\
            \frac{|E(i,j)|}{|E(j)|} & \quad \text{otherwise}\\
          \end{array} \right.
        \end{equation}

         In the above equation allowing casting two votes instead of one accommodates for the situations where a genuine "mind change" has taken place, as explained earlier. The spamicity parameter is used in Section \ref{sec:Ind} as an indicator for finding collusion groups.
 \item
        The time format in AMZLog is `yyyy-mm-dd'. When preprocessing data, we change this time to an integer number. The number shows the number of days between the date on which the first ranking score in the system has been posted and the day in which the rating score has been cast.
\end{itemize}

\section{Finding Collusion biclique Candidates} \label{sec:cluster}

In order to find collusion bicliques, the first step is to identify all collaborative rating groups as collusion biclique candidates, then check them to find real collusion groups. We use Algorithm \ref{alg:allclq} which employs FIM technique to find bicliques. Algorithm \ref{alg:allclq} is applied to an initialized array of biclique candidates denoted by $C$ and gives the list of all bicliques denoted by $FC$. To initialize $C$, we generated a corresponding biclique for every reviewer. The biclique contains the ID of the reviewer as the only member of its $R$, the list of all products the reviewer has rated as the members of $P$ and all rating scores cast by the reviewer as the members of $Rel$. The $NextC$ is a temporary list of bicliques used just for transferring groups found in every iteration to the next iteration of the algorithm.

In every step we firstly combine all members of $C$, to find possible larger groups. The problem of finding all collaboration groups is an NP-Complete problem. We add some constrains to decrease the time complexity and solve the problem in an accepted time period. These constrains are mainly applied as the accepted size of the $R$ and $P$ for every group. We suppose that number of reviewers in every group should be at least $MinR$ and the number of products at least $MinP$. We find the optimum values of $MinR$ and $MinP$ by experiment. $MinR$ and $MinP$ may impact the number of identified groups, but will not affect the process of collusion detection. We use the minimum possible values for these variables which are $2$ for $MinR$ and $3$ for $MinP$. We do n set $MinP$ to $2$, because it is highly possible that two honest reviewers rate two same products by accident, similarity of interest, etc.

\begin{algorithm}[!t]
\caption{Finding All biclique candidates}
\label{alg:allclq}
\textbf{Input:} $C$ as the set of all initial bicliques. \\
\textbf{Output:} $FC$ as the set of all candidate spam bicliques\\
\begin{algorithmic}
    \REPEAT
        \STATE Empty $NextC$
        \FORALL{$c \in C$}
            \STATE $c.Processed=false$
        \ENDFOR
        \FOR{$i = 1 \to (C.size -1)$}
            \FOR{$j = i+1 \to C.size $}
                \STATE $c3.P = c_i.P \cap c_i.P$ /* $c3$ is a temporary variable */
                \IF {$c3.P.size \geq MinP$}
                    \STATE $c3.R = c_i.R \cup c_j.R$
                    \STATE Add $c3$ to $NextC$
                    \STATE $c_i.Processed = true$
                    \STATE $c_j.Processed = true$
                \ENDIF
            \ENDFOR
        \ENDFOR
        \FORALL{$c \in C$}
            \IF {$(c.Processed == false \wedge (c.R.size \geq MinR))$}
                \STATE Move $c$ To $FC$
            \ENDIF
            \STATE Remove $c$ from $C$

        \ENDFOR
        \STATE $C$ = $NextC$
        \STATE Empty $NextC$
    \UNTIL{$(C.size> 0)$}
    \RETURN $FC$
\end{algorithmic}
\end{algorithm}

\section{Indicators} \label{sec:Ind}
\subsection{Collusion Indicators}
It is almost impossible to find collusion bicliques by analyzing the behavior of individual members or even only on the posted rating scores. Rather, several indicators must be checked to realize in what extent a group is a collusion group \cite{mainWWW}. In this section we propose some indicators and show how they indicate to a possible collusive activity.

\subsubsection{Group Rating Value Similarity (GVS)}
A group of reviewers can be considered as a collusion biclique, if they are posting similar rating values for similar products. For example, all of them promote product $p_1$ and demote $p_2$ by similar rating scores. To check this similarity we at first find the \emph{Pairwise Rating Value Similarity} between every pair of reviewers in the group. Pairwise rating value similarity denoted by $VS(i,j)$ shows in what extent $u_i$ and $u_j$ have cast similar rating scores to every product in $g.P$.

We use cosine similarity model, a well-known model for similarity detection \cite{cosine}, for finding similarities between group members. Since $VS(i,j)$ is the cosine of the angle between two vectors containing ratings of two reviewers, it is a value in range $[0,1]$. The value $1$ means completely same and $0$ means completely different.

Suppose that $u_i$ and $u_j$ are two members of group $g$ i.e. $u_i$ and $u_j \in g.R$ . We calculate similarity between them as follows:

\begin{equation}
\label{eq:VS}
VS(i,j) = \frac{\sum_{k \in g.R} v_{ik} \times{} v_{jk}}{\sqrt{\sum_{k \in g.R}(v_{ik})^2}  \times{} \sqrt{\sum_{k \in g.R}(v_{jk})^2}}
\end{equation}
Then we calculate an overall degree od similarity for every group to show how all members are similar in terms of value they have posted as rating scores and call it Group Rating Value Similarity (GVS). GVS for every group is the minimum amount of pairwise similarity between group members.So, the GVS of group $g$ is calculated as follows:

\begin{equation}
\label{eq:GVS}
GVS(g) = \min (VS(i,j)), \text{ for all } u_i \text{ and } u_j \in g.R
\end{equation}

The bigger the $GVS$, the more similar the group members are in terms of their posted rating values.

\subsubsection{Group Rating Time Similarity (GTS)}

Another indicator for finding collusive groups is that they usually cast their ratings in a short period of time. In other words, when a group of users try to promote or demote a product they attempt to do it in a short period of time to gain benefits, get paid, etc. We can use this as an indicator to find collusion groups. To find calculate Group rating Time Similarity (GTS), we at first find the the Time Window (TW) in which the rating scores have been posted on product $p, p\in g.P$. The time window starts when the first rating posted and ends when the last one cast. We use the size of the time window in comparison to a constant $MaxTW$ to show how big this time window is. The parameter $MaxTW$ is the maximum size of a time window which my be supposed as collusive. We suppose that the time windows larger than $MaxTW$ are wide enough and we do not suppose them az indicators to a possible collusion. So for every product in the group we have:
\begin{equation}
\label{eq:TW}
  TW(j)  = \left\{
  \begin{array}{l}
    0 \quad \text{, if $MaxT(j)- MinT(j) > MaxTW$} \\ \\
    1- \frac{MaxT(j) - MinT(j)}{MaxTW}  \quad \text{ , Otherwise}\\
  \end{array} \right.
\end{equation}
where

\begin{equation}
\begin{split}
\nonumber
MinT = \min(e_{ij}.t) \text{ , }\\
MaxT = \max(e_{ij}.t) \text{ , }\\
j \in g.P \text{ for all } i \in g.R
\end{split}
\end{equation}

Now, we choose the largest $TW(j)$ as the degree of time similarity between the ratings posted by group members on the target products. Therefore we say:
\begin{equation}
\label{eq:GTS}
GTS(g) = \max (TW(j)), \\ \text{ for all } p_j \in g.P
\end{equation}

The bigger the $GTS$, the more similar the group members are in terms of their posted rating times.

\subsubsection{Group Ratings Spamicity (GRS)}

As we described in Section \ref{ssec:pp}, the spamicity of a rating shows the suspiciousness of the rating score because of high number of ratings posted by same user to a same product ( equation~(\ref{eq:spam})). We define the Group Rating Spamicity (GRS) as the ratio of the weight of spam ratings and the weight of all ratings in the group. We calculate GRS as follows:
\begin{equation}
\label{eq:GRS}
GRS(g) = \frac{\sum_{e \in g.Rel} e.v \times{} e.spam }{\sum_{e \in g.Rel} e.v}
\end{equation}

\subsubsection{Group Members Suspiciousness (GMS)}
Suspicious users are users whose behavior indicates that they can be potentially colluders. We identify suspicious users in four steps.

\noindent \textbf{Step 1:} Suppose that $E(j)$ is the set of all ratings posted for the product $p_j$. We find the median of the members of $E(j)$ and denote it by $m_j$. Then, we calculate the average distance of all ratings posted on $p_j$ from the median $m_j$ and denote it by $d_j$. We calculate the average distance using equation (\ref{eq:s}).

\begin{equation}
\label{eq:s}
d_j = \sqrt{ \frac{\sum_{i \in E(j)} (e_{ij}.v - m_j) ^2 }{||E(j)||}}
\end{equation}

Now, we calculate a credibility degree for every $e_{ij} \in E$. We denote this credibility degree $\varphi_{ij}$ and use it to eliminate the ratings fall far away from the majority. The $\varphi_{ij}$ is calculated as follows.

\begin{equation}
\label{eq:crd}
  \varphi_{ij} = \left\{
  \begin{array}{l l}
    1 & \quad \text{if $(m_j - d_j) \leq r_{ij} \leq (m_j + d_j)$}\\
    0 & \quad \text{otherwise}\\
  \end{array} \right.
\end{equation}

Equation (\ref{eq:crd}) shows that the ratings which fall in range $m_j \pm d_j)$ are considered as credible and the rest are identified as unfair.

\noindent \textbf{Step 2:} In this step, we recalculate the averages of all credible ratings on the product $p_j$ and assume that it is a dependable guess for the real rating of the product. We show it by $g_j$ and calculate it as follows.

\begin{equation}
\label{eq:g}
g_j = \frac{ \sum_{i \in E(j)} (e_{ij}.v \times{} \varphi_{ij})}{\sum_{i \in E(j)} \varphi_{ij}}
\end{equation}

\noindent \textbf{Step 3:} In the third step, using the ratings we calculated for every product we build two error rates for every reviewer. The first error rate is the \emph{$L^p$ error rate} which is  $L^p-norm$ of all differences between the ratings cast by the reviewer on the quality of $p_j$ and the $g_j$. We denote the $L^p$ error rate of reviewer $u_i$ by $LP(i)$. Suppose that $J_i$ is the set of indices of all products have been rated by $u_i$. The $LP(i)$ we calculate is in fact $L^2-norm$ and is calculated as follows.

\begin{equation}
\label{eq:L}
LP(i) =   \sqrt{\sum_{j \in J_i} \Big( \Big| e_{ij}.v - g_j \Big| \Big)^2}
\end{equation}

The second error rate we calculate for every reviewer is the uniform norm of differences between $e_{ij}.v$ and $g_j$ for all products have been rated by $u_i$. We call this error rate \emph{uniform error rate }of $u_i$, denote it by $UN(i)$ and calculate as follows:

\begin{equation}
\label{eq:U}
UN(i) = \max \Big(\Big|e_{ij}.v - g_j \Big|\Big) \quad ,\quad j \in J_i
\end{equation}

\noindent \textbf{Step 4:} In this step, we trap the suspicious reviewers. The suspicious reviewers are the people who have have large $LP(i)$ or while they have normal $LP(i)$ they have high $UN(i)$ values.  To identify suspicious reviewers, we try to identify the normal range of error rates for all reviewers. Based on the calculated range, the outliers are considered as suspicious reviewers. Suppose that $\widehat{LP}$ is the median of all $LP(i)$ and $\widehat{UN}$ is the median of all $UN(i)$. Also, we assume that $\overline{LP}$ and $\overline{UN}$ are the standard distance of all $LP(i)$ and $UN(i)$ respectively , calculated similar to the method proposed in equation (\ref{eq:s}). The list of suspicious reviewers is denoted by $S$ and built using equation \ref{eq:S}.

\begin{equation}
\label{eq:S}
\begin{split}
S = \Big\{ u | \Big(u \in U \Big) \text{ and } \Big( \big(LP(u)\text{ in } (\widehat{LP} \pm \overline{LP})\big) \\
\text{ or } \\
 \big(UN(u) \text{ in } (\widehat{UN} \pm \overline{UN})\big)\Big) \Big\}
\end{split}
\end{equation}


Now, we define GMS of a group as the ratio of the suspicious members of a group and the total number of members of a group. Hence, suppose that for group $g$, $S(g)$ is the list of suspicious users of the group. The GMS is calculated as follows:

\begin{equation}
\label{eq:S}
GMS(g) = \frac{|S(g)|}{|g.R|} \text{ where } S(g) = (g.R) \cap S
\end{equation}

\subsection{Defectiveness Indicators}

The defectiveness of a collusion group is its power to impact normal behavior of system and the ratings of products. There are two important parameters reflecting defectiveness of a group: size of the group and the number of products have been attacked by the group. The bigger these two parameters are, the more defective the collusive group is.

\subsubsection{Group Size (GS)}
Size of a collusive group (GS) is proportional to the number of reviewers who have collaborated in the group ($g.R.size$). The GS is calculated using the equation (\ref{eq:GS}).
\begin{equation}
\label{eq:GS}
GS(g) = \frac{|g.R|}{\max(|g.R|)} \text{ where } g \text{ is a group}
\end{equation}
GS is a parameter between $(0,1]$ and showing how large is the number of members of a group in comparison with other groups.

\subsubsection{Group Target Products Size (GPS)}
The size of the target of a group (GPS) is proportional  to the number of products which have been attacked by members of a collusion group. The bigger the GTS is, the more defective the group will be. GTS is calculated as follows:
\begin{equation}
\label{eq:GTS}
GPS(g) = \frac{|g.P|}{\max(|g.P|)} \text{ where } g \text{ is a group}
\end{equation}
GTS as proposed in equation \ref{eq:GTS}, is a number in range $(0,1]$ showing how large is the size of the group target in comparison with other groups.

\section{Finding Collusive Sub-Bicliques} \label{sec:sub}

When finding bicliques, we try to maximize size of the group, i.e. $g.R.size$, to find the largest possible candidates bicliques. It is possible that a large group be identified as an honest group due to large and diverse number of users who have just rated same sub-set of products. But possibly, there exist some smaller groups inside the large groups that can build a collusive sub-groups. In this section we propose an algorithm for finding these possible sub-groups.

Algorithm \ref{alg:allsclq} shows the process of finding sub-groups in a candidate group $g$. In this algorithm, we use the idea of finding candidate bicliques in Section \ref{sec:cluster}. At first we build one biclique for every relation in the $g.Rel$. Then we try to merge every two generated bicliques. If the resulted biclique has both $GVS$ and $GTS$ greater than a threshold called $\delta$, we keep it for next steps. we continue this process until all possible sub-groups are identified. The reason that $GVS$ and $GTS$ are chosen is that if a group does not have similar rating scores or they are posted in a wide time range there is a very small possibility that the group is a collusive group \cite{mainWWW}. Threshold $\delta$ shows the minimum value of DOC and DI and its default value experimentally is set to $0.4$ (see Section \ref{sec:eval}). Users can change the value of $\delta$ as they intend. We use $\delta$ for finding sub-bicliques and also querying bicliques \ref{sec:frm}.

\begin{algorithm}[!t]
\caption{Finding Possible Collusive Sub-Bicliques}
\label{alg:allsclq}
\textbf{Input:} $cl$ as a biclique. \\
\textbf{Output:} $Scl$ as the set of all collusive sub-bicliques, initially empty\\
\begin{algorithmic}
    \STATE Build $C$ as the set of sub-bicliques of $cl$ each having only one relation.
    \STATE Empty $NextS$
    \REPEAT
        \FORALL{$c \in C$}
            \STATE $c.Processed=false$
        \ENDFOR
        \FOR{$i = 1 \to (C.size -1)$}
            \FOR{$j = i+1 \to C.size $}
                \STATE $c_3.P =$ Merge($ c_i.P, c_i.P$) /* $c_3$ is a temporary variable */
                \STATE Calculate all indicators for $c_3$
                \IF {$(c.DI > \delta )$}
                    \STATE Add $c_3$ to $NextS$
                    \STATE $c_i.Processed = true$
                    \STATE $c_j.Processed = true$
                \ENDIF
            \ENDFOR
        \ENDFOR
        \FORALL{$c \in C$}
            \IF {$(c.Processed == false) \wedge (c.P.size > MinP) \wedge (c.R.size > MinR)$}
                    \STATE Move $c$ To $Scl$.
                \STATE Remove $c$ from $C$
            \ENDIF
        \ENDFOR
        \STATE $C$ = $NextS$
        \STATE Empty $NextS$
    \UNTIL{$(C.size> 0)$}
    \RETURN $Scl$
\end{algorithmic}
\end{algorithm}
$\delta$ is the collusion threshold. Details can be found in Section~\ref{sec:frm}.
\section{Collusion Detection Framework} \label{sec:frm}
In this section we propose a collusion detection framework. We discuss the basics and principals of the framework, its overall architecture and the query language we propose in thispaper.for and show how it can help users query collusion graph.
\subsection{Overview}
To find collusive groups we firstly calculate collusion indicators for every identified biclique. Our proposed collusion detection model employs algorithms (\ref{alg:allclq}) and (\ref{alg:allsclq}) and indicators proposed in section \ref{sec:Ind} to effectively find collusive bicliques. It is notable that it is not possible to automatically find out if a group is collusive or not. So, using the calculated indicators we build a \emph{degree of collusion} (DOC) for every biclique. DOC shows in what extent a group behaves like a collusion group. DOC is an aggregation of four collusion indicator. Since in different environments, the importance of these indicators may be different, our models enables users to assign weight to every indicator to have its adequate weight in DOC. Suppose that $W_{GVS}$, $W_{GTS}$, $W_{GRS}$ and  $W_{GMS}$ are corresponding weights for $GRS$, $GTS$, $GRS$ and $GMS$ respectively so that $W_{GVS} + W_{GTS}+ W_{GRS}+ W_{GMS} =1$. The default values of all these weights are $0.25$. The DOC is calculated as follows:

\begin{equation}
\label{eq:GS}
\begin{split}
DOC(g) = GVS\times{}W_{GVS}+ GTS \times{}W_{GTS} \\
 + GRS \times{} W_{GRS}+ GMS \times{} W_{GMS}
\end{split}
\end{equation}

Moreover, for every group we calculate damaging impact (DI) to show how damaging the group can be in the system. Two parameters are used in calculating DI. The first parameter is $GPS$, that shows the ratio of the total products which have been targeted by the collusion group and $GS$ which shows the ratio of the users which have been included in collusion. We calculate DI as follows:

\begin{equation}
\label{eq:DI}
DI(g) = \frac{GPS+GTS}{2}
\end{equation}

Users is also able to specify the threshold $\delta$ for collusion.

The process showing how we find bicliques, sub-bicliques and calculate their related indicators are proposed in Algorithm \ref{alg:final}.

\begin{algorithm}[!t]
\caption{Collusion Detection Algorithm }
\label{alg:final}
\textbf{Input:} $ORM$ Graph. \\
\textbf{Output:} $C$ as the set of all collusive Bicliques, initially empty\\
\begin{algorithmic}
    \STATE  $TempC = $ All Bicliques //Use ALG.\ref{alg:allclq}
    \FORALL{$tc \in TempC$}
        \STATE Calculate Indicators for $tc$
    \ENDFOR
    \REPEAT
        \STATE $tc = $ first member of $TempC$
        \IF {$tc.DOC > \delta$}
            \STATE Move $tc$ to $C$
        \ELSE
            \IF {$tc.DI < \delta $}
                \STATE Remove $tc$ from $TempC$
            \ELSE
                    \STATE $SC =$ Sub-Bicliques of $tc$ // Use ALG.\ref{alg:allsclq}
                    \STATE Calculate Indicators for all members of $SC$
                    \STATE $TempC = TempC \cup SC $
                    \STATE Remove $tc$ from $TempC$
            \ENDIF
        \ENDIF

    \UNTIL{$(TC.size> 0)$}
    \RETURN $C$
\end{algorithmic}
\end{algorithm}

\begin{figure*}[!t]
\centering
  \includegraphics[scale=0.5]{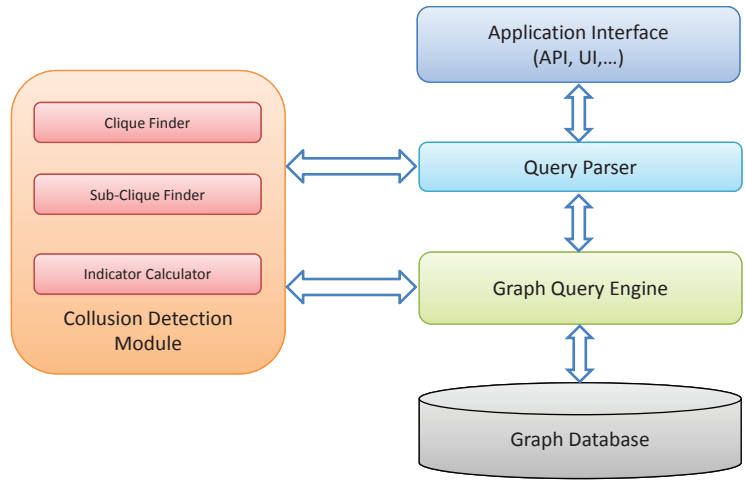}\\
  \caption{Overall Architecture of Proposed Model.}\label{arch}
\end{figure*}

\subsection{Architecture}
To detect collusion bicliques and enable people to easily query collusion graph, our model comprises an architecture consists of several parts. The architecture is proposed in Figure \ref{arch}. The followings are description of what every parts do in the architecture.
\begin{itemize}
  \item
        \textbf{Collusion Detection Part} as the main part of the model is responsible for finding bicliques. It is also contains the indicator calculator which is responsible for calculating collusion indicators corresponding to every detected biclique. Moreover, this part investigate large non-collusive bicliques to find any possible collusive sub-bicliques. Algorithm \ref{alg:final} shows responsibilities of this part.
  \item
        \textbf{Application Interface} is the part which is the means which enables users to query the ORM graph.  The queries can be generated by a query builder provided as a user interface or they can be posted to the system via API.
  \item \textbf{Query Parser} is responsible for parsing queries received from application interface. It parser queries and converts them to a SPARQL query. By customizing this parser, it is possible to put this framework on any database management system.
  \item \textbf{Graph DBMS} executes queries receives from query parser and returns the results to it. This engine should be chosen in accordance with the format of queries received from query parser.
  \item \textbf{Database} is a triple-store database. We use RDF triples for representing data and store them in an RDF store. So, any storage system capable of saving and retrieving data in RDF format can be used as this part.
\end{itemize}

\subsection{Query Language}
In our model, online rating systems are represented as a graph, so it can be processed and queried using existing graph query languages like SPARQL. The overall format of a query is as follows:

\begin{alltt}
getbicliques[.product|.reviewer]([\(v\),\(t\),\(r\),\(m\)])
[
filter\{
       [on(list of products);]
       [contain(list of users);]
       [DOC > delta;]
       \}
       ];
\end{alltt}

This query enables users to:

\begin{itemize}
  \item
        Look for specific bicliques containing one or more particular reviewers in their $R$ or products in their $P$.
  \item
        Look for products which have been attacked by one or more specific reviewers.
  \item
        Look for reviewers who have been collaborated in attacking to one or more particular products.
  \item
        Specify $\delta$ as the minimum level of collusiveness in all queries.
  \item
        Specify weights for collusion indicators to build more customized queries and get more adequate results. $v$, $t$, $r$ and $m$ are corresponding weights for $GRS$, $GTS$, $GRS$ and $GMS$ respectively.
\end{itemize}

The followings are sample queries that show the usability of the query language:

\addtocounter{dc}{1}
\noindent \emph{Example \arabic{dc}.} Adam , a user, wants to find all collusion bicliques in the system. He uses all default vales, so his query will be as:
\begin{alltt}
    getbicliques();
\end{alltt}

\addtocounter{dc}{1}
\noindent \emph{Example \arabic{dc}.} Adam wants to find all collusion bicliques in the system. He wants to see serious attacks so he prefers to see bicliques with $DoC > 0.7$. So he should write a query like this:
\begin{alltt}
    getbicliques()
    filter\{
        DOC > 0.7;
    \};
\end{alltt}

\addtocounter{dc}{1}
\noindent \emph{Example \arabic{dc}.} Adam is administrator of a system in which similarity of values is so more important twice time than others. So, he prefers not to use default weights. Rather, he specifies weights by himself and designs the query as follows:
\begin{alltt}
    getbicliques(0.4,0.2,0.2,0.2);
\end{alltt}

\addtocounter{dc}{1}
\noindent \emph{Example \arabic{dc}.} With this new indicators weights, Adam intends to see all products which have been attacked by groups contain `Jack' and `Jhon'. So he designs the query as follows:
\begin{alltt}
    getbicliques.products(0.4,0.2,0.2,0.2);
    filter\{
        contains(`Jack', `Jhon');
    \};
\end{alltt}

\addtocounter{dc}{1}
\noindent \emph{Example \arabic{dc}.} Adam intends to see all users who have been collaborated on unfairly ranking products `Book1' and `DVD2'. So he may design the query as follows:
\begin{alltt}
    getbicliques.reviewers(0.4,0.2,0.2,0.2);
    filter\{
        on(`Book1',`DVD2');
    \};
\end{alltt}

\addtocounter{dc}{1}
\noindent \emph{Example \arabic{dc}.} As a more detailed query, Adam intends to see all bicliques containing `Jack' and `Jhon' and their targets contain `Jack' and `Jhon'. Meanwhile he wants to see serious attacks with $DOC > 0.7$ ad indicators weights as $0.4$,$0.3$,$0.2$ and $0.1$. He may design the query as follows:
\begin{alltt}
    getbicliques(0.4,0.3,0.2,0.1)
    filter\{
        contains(`Jack', `Jhon');
        on(`Book1',`DVD2');
        DOC > 0.7;
    \};
\end{alltt}


\section{Implementation and Evaluation} \label{sec:eval}
\subsection{Implementation}
To implement our framework we used several techniques for various part of the system. As graph query language we can use any variant of SPARQL like query languages. As we have several bicliques to manage here, which every biclique is in fact a subgraph, we should use a version of SPARQL which simplifies using and managing sub-graphs. We use FPSPARQL, a folder-enabled query language proposed in our previous work  \cite{BPM11} which makes dealing with bicliques much easier. We also have used engine designed for FPSPARQL as Graph DBMS. The performance of this query engine is tested and verified in \cite{BPM11}.

We also have implemented a front-end tool to assist users using our query language. The tool which is shown in Figure \ref{impl}, provides a friendly easy to use interface with a set of pre-designed templates which makes writing queries easier. Users can choose a query template and customize it to fit their needs or they can design a query from scratch. Then they can run the query and see the results on the screen. The query language is also available via API. It means that other applications can easily contact the tool and query collusion graph when they need. The snapshot shows the results received by running following query:
\begin{alltt}
    getbicliques(0.4,0.3,0.2,0.1)
    filter\{
        contains('u1','u2','u3','u6','u7');
        on('p1','p2','p3','p5','p6','p7');
        DOC > 0.7;
    \};
\end{alltt}
\begin{figure}[!t]
\centering
  \includegraphics[scale=0.22]{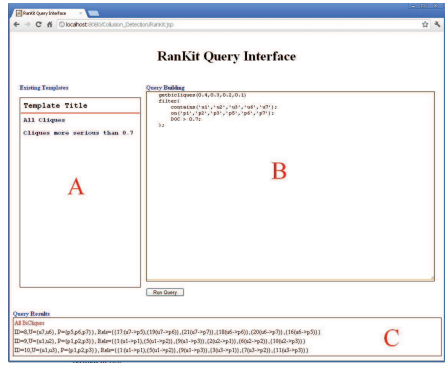}\\
  \caption{Snapshot of our front-end tool interface, (A) Predefined Query Templates, (B) Query Builder and (C) Results Pane.}\label{impl}
\end{figure}

\begin{figure}[!t]
\centering
  \includegraphics[scale=0.5]{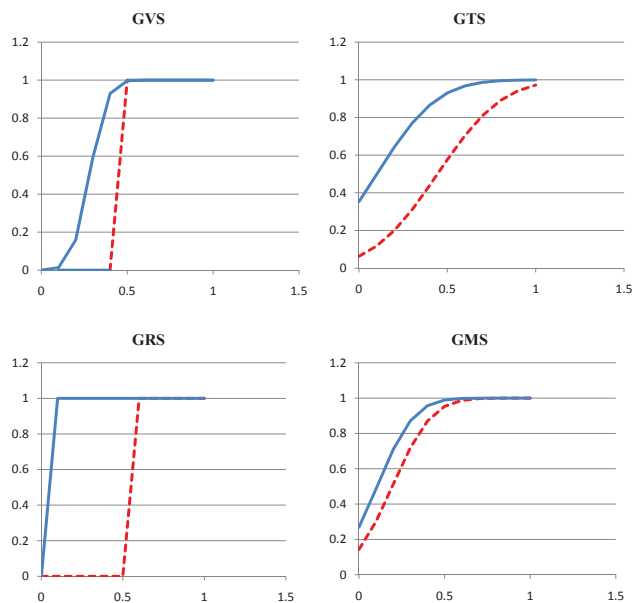}\\
  \caption{Cumulative \% of values of collusion indicators for collusive(red/dashed) and non-collusive (blue/solid) selected groups.}\label{fig:Stats}
\end{figure}

\subsection{Experimentation and Evaluation}
To evaluate our model, we randomly chose $20$ groups identified within AMZLog. Then we asked domain experts to manually check whether these groups are collusive or not. Experts using their experience and information provided in Amazon web site such as rating scores, corresponding written reviews,etc analyzed these $20$ selected groups. The result was $7$ out of $20$ groups were collusive and $13$ groups were identified as honest groups.We used this dataset to evaluate our model.

Our model contains three main parts. The first part is the algorithms proposed for finding bicliques and sub-bicliques. These algorithms follow the FIM model. The performance evaluation of FIM is proposed in its original paper  \cite{FIM} and its is being used in other similar works like \cite{mainWWW}.

The second part is the query language proposed for querying collusion graph. All designed queries are directly converted to FPSPARQL queries using query parser and then are run on its FPSPARQL query engine. The performance of the FPSPARQL query engine is tested, and results are available in \cite{BPM11}.

The third important part of our model are the indicators which are proposed for collusion detection and the quality of results they provide. We evaluate this aspect of our model in following subsections.

\subsection{Statistical Evaluation}
We calculate four collusion indicators for collusive groups . Then we calculate cumulative distribution of values calculated for every indicator for all groups. We do the same process also for non-collusive groups. The results of these calculations are shown in Figure \ref{fig:Stats}. In every part of the figure, the vertical axis is the cumulative value and horizontal axis is the percentage.
For every indicator in figure \ref{fig:Stats}, the cumulative distribution chart of the non-collusive groups are on the left side while collusive charts are more closer to one. It simply means that in average, for every indicator, values calculated for collusive groups are larger than calculated values for non-collusive groups.  Therefor, these indicators and the way they are calculated are truly reflect the behavior of the group, can be used as indicators to identify collusion groups.

\subsection{Evaluating Quality of the Results}
To evaluate quality of the results returned by our model, we use the well-known measures of precision and recall \cite{precision}. Precision measures the quality of the results and is defined by the ration of the relevant results to the total number of retrieved results. Recall measures coverage of the relevant results and is defined by the ratio of relevant results retrieved to the total number of existing relevant results in database. With respect to collusion detection, we define precision and recall in equations (\ref{eq:prec}) and (\ref{eq:recall});

\begin{equation}
\label{eq:prec}
\begin{split}
precision = \frac{\text{\scriptsize{Number of retrieved bicliques which are really collusive}}}{\text{\scriptsize{total number of retrieved bicliques}}}
\end{split}
\end{equation}

\begin{equation}
\label{eq:recall}
\begin{split}
Recall = \frac{\text{\scriptsize{Number of retrieved bicliques which are really collusive}}}{\text{\scriptsize{total number of collusive bicliques in database}}}
\end{split}
\end{equation}

\begin{figure}[!t]
\centering
  \includegraphics[scale=0.6]{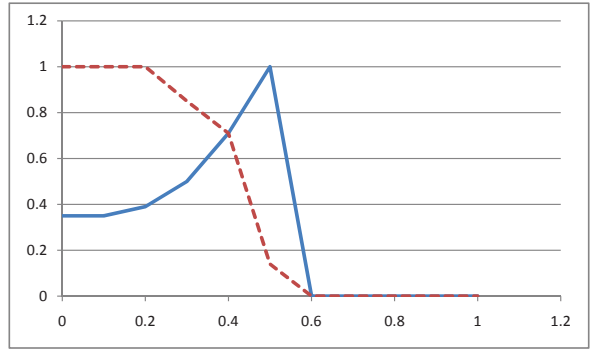}\\
  \caption{How precision (blue/solid) and recall (red/dashed) change with different threshold values.}\label{fig:pr}
\end{figure}

An effective model should achieve a high precision and a high recall. But its is not possible in real world, because these metrics are inversely related \cite{precision}. It means that cost of improvement in precision is reduction of recall and vice versa. We calculate precision and recall with different thresholds to show how changing impacts on the quality and accuracy of the results. Figure \ref{fig:pr} shows the results of running model with different threshold values. We do not specify particular value for precision and recall. We can say that if the user wants to achieve the highest possible values for both precision and recall metrics, figure \ref{fig:pr} obviously shows that the optimal value for threshold ($\gamma$) is $0.4$. In this case 71\% of the bicliques are retrieved and 71\% of retrieved results are really collusive. Using query language, user has opportunity to increase quality or coverage of data by changing threshold value.

As we described before, our model also calculates damaging impact (DI) of every group to show how damaging is the group. DI helps users to identify groups that have a high potential power to damage the system. These groups are mainly groups with high number of targeted products. Table~\ref{tbl:di} shows a sample list of bicliques and their DI and DOC. The group $2$ has high values for DOC and DI, so it is a dangerous group. On the other hand, group $9$ has mall values for DOC and DI, and is not a really dangerous group. Looking at group $11$ reveals that although the DOC of the groups is small (0.121), but it still can be divided as dangerous because its DI is 0.41 showing that the damaging potential of the group is still high. DI is also useful when comparing groups $4$ and $6$. Whereas, the DOC of group $6$ is $8\%$ higher than DOC of group $4$, but the DI of group $4$ is $7$\% higher than DI of group $6$. Therefore we can judge them similar rather than putting group $4$ after group $6$. Without having DI, the damaging potential of groups like $11$ or $4$ may be underestimated. This may lead to unfairly manipulating rating scores.

\begin{table}
    \begin{center}
       \rowcolors{1}{lightgray}{green}
        \begin{tabular}{ |l |c | r|r| }
            \hline
             Group ID & Degree of Collusiveness & Damaging Impact \\
             \hiderowcolors
            \hline
              2 & 0.598 & 0.599 \\
            \hline
              4 & 0.37 & 0.49 \\
            \hline
              6 & 0.45 & 0.42 \\
            \hline
              9 & 0.056 & 0.173 \\
            \hline
              11 & 0.121 & 0.41 \\
            \hline
        \end{tabular}
  \end{center}
  \caption{Sample Bicliques with their Degree of Collusiveness and Damaging Impact.}
  \label{tbl:di}
\end{table}

\section{Related Work} \label{sec:rel}
Collusion detection has been widely studied in P2P systems \cite{CollusionInP2P,eigentrust,EigenCollusion}. For example EigenTrust \cite{eigentrust} tries to build a robust reputation score for p2p collaborators but a research \cite{EigenCollusion} show that it is still subject to collusion. A comprehensive survey on collusion detection in P2P systems can be found here \cite{CollusionInP2P}. Models proposed for detecting colluders in P2P systems are not applicable to online rating systems because in P2P systems, models are mostly built on relations and communications between people. But in online rating systems there is no direct relation between raters.

Reputation management systems are also targeted by collusion\cite{IEEE2102Survey,reliable}. Very similar to rating systems, colluders in reputation management systems try to promote or demote reputation  scores by Collusion. Some work try to identify collusion using majority rule \cite{betarep,maj2}. The majority rule is prone to attacks which hijack the product or colluders try to manipulate reputation slowly by posting scores not far from majority of community. The other method is weighting votes based on the reputation of the voter \cite{defending,weight1,pagerank}. Temporal analysis of the behavior of the voters is another way for collusion detection \cite{temporal} which is not sufficient without analyzing other aspects of voting like reputation of the voter or value of the vote. Yang et. al. \cite{reptrap} try to  identify collusion by employing both majority rule and temporal behavior analysis. Their model is not still tested thoroughly and just is applied to a specific dataset and a particular type of attack.

The most similar work to ours, is \cite{mainWWW}. In this work Mukherjee et.al. propose a model for spotting fake review groups in online rating systems. The model analyzes textual feedbacks cast on products in Amazon online market to find collusion groups. They use $8$ indicators to identify colluders and propose an algorithm for ranking collusion groups based on their degree of spamicity. However, our model is different from this model in terms of proposed indicators, analyzing personal behavior of the raters and dealing with redundant rating scores. Also a recent survey \cite{ecomsurvey} shows that buyers rely more on scores and ratings when intend to buy something rather than reading textual items. So, in contrast with this model we focus on numerical aspect of posted feedback. However, the model proposed by Mukherjee et.al. is still vulnerable to some attacks. For example, if the number of attackers is much higher than honest raters on a product the model can not identify it as a potential case of collusion.

Another major difference between our work and other related work is that, we propose a graph data model and a also flexible query language for better understanding, analyzing and querying collusion. This aspect is missing in almost all previous work.

\section{Discussion and Conclusion} \label{sec:Concl}
In this paper we proposed a novel framework for collusion detection in online rating systems. We used two algorithms designed using frequent itemset mining technique for finding candidate collusive groups and sub-groups in our dataste. We propose several indicators showing the possibility of collusion in group from different aspects. We used these indicators to assign every group a rank to show their degree of collusiveness and also damaging impact. We also propose a query language and also a front-end tool to assist users find collusion groups according their intended criteria. We evaluated our model first statically and showed the adequacy of the way we define and calculate collusion indicators. Then we used precision and recall metrics to show quality of output of our model.

As future direction, we plan to identify more possible collusion indicators. We also plan to extend the proposed query language with more features and en enhanced visual query builder to assist users employing our model. Moreover, we plan to generalize our model apply it to other possible areas which are subject to collusive activities.

%
\bibliographystyle{abbrv}
\bibliography{CD_TechRep.bbl}

\begin{thebibliography}{10}

\bibitem{FIM}
R.~Agrawal and R.~Srikant.
\newblock Fast algorithms for mining association rules in large databases.
\newblock In {\em Proceedings of the 20th International Conference on Very
  Large Data Bases}, VLDB '94, pages 487--499, San Francisco, CA, USA, 1994.
  Morgan Kaufmann Publishers Inc.

\bibitem{BPM11}
S.-M.-R. Beheshti, B.~Benatallah, H.~R.~M. Nezhad, and S.~Sakr.
\newblock A query language for analyzing business processes execution.
\newblock In {\em BPM}, pages 281--297, 2011.

\bibitem{CollusionInP2P}
G.~Ciccarelli and R.~L. Cigno.
\newblock Collusion in peer-to-peer systems.
\newblock {\em Computer Networks}, 55(15):3517 -- 3532, 2011.

\bibitem{cswww}
A.~Doan, R.~Ramakrishnan, and A.~Y. Halevy.
\newblock Crowdsourcing systems on the world-wide web.
\newblock {\em Commun. ACM}, 54:86--96, April 2011.

\bibitem{ecomsurvey}
A.~Flanagin, M.~Metzger, R.~Pure, and A.~Markov.
\newblock User-generated ratings and the evaluation of credibility and product
  quality in ecommerce transactions.
\newblock In {\em System Sciences (HICSS), 2011 44th Hawaii International
  Conference on}, pages 1--10. IEEE, 2011.

\bibitem{amazonproblem}
A.~HARMON.
\newblock Amazon glitch unmasks war of reviewers.
\newblock In {\em NY Times (2004, Feb. 14)}.

\bibitem{ebayProblem}
J.~M. J.~Brown.
\newblock Reputation in online auctions: The market for trust.
\newblock {\em CALIFORNIA MANAGEMENT REVIEW}, 49(1):61 --81, Fall 2006.

\bibitem{maj2}
W.~Jianshu, M.~Chunyan, and G.~Angela.
\newblock An entropy-based approach to protecting rating systems from unfair
  testimonies.
\newblock {\em IEICE TRANSACTIONS on Information and Systems},
  89(9):2502--2511, 2006.

\bibitem{betarep}
A.~Jsang and R.~Ismail.
\newblock The beta reputation system.
\newblock In {\em Proceedings of the 15th Bled Electronic Commerce Conference},
  pages 41--55, 2002.

\bibitem{eigentrust}
S.~D. Kamvar, M.~T. Schlosser, and H.~Garcia-Molina.
\newblock The eigentrust algorithm for reputation management in p2p networks.
\newblock In {\em Proceedings of the 12th international conference on World
  Wide Web}, WWW '03, pages 640--651, New York, NY, USA, 2003. ACM.

\bibitem{thesis}
R.~Kerr.
\newblock Coalition detection and identification.
\newblock In {\em Proceedings of the 9th International Conference on Autonomous
  Agents and Multiagent Systems: volume 1 - Volume 1}, AAMAS '10, pages
  1657--1658, Richland, SC, 2010. International Foundation for Autonomous
  Agents and Multiagent Systems.

\bibitem{temporal}
P.~Laureti, L.~Moret, Y.~Zhang, and Y.~Yu.
\newblock Information filtering via iterative refinement.
\newblock {\em EPL (Europhysics Letters)}, 75:1006, 2006.

\bibitem{simplified}
H.~Lee, J.~Kim, and K.~Shin.
\newblock Simplified clique detection for collusion-resistant reputation
  management scheme in p2p networks.
\newblock In {\em Communications and Information Technologies (ISCIT), 2010
  International Symposium on}, pages 273 --278, oct. 2010.

\bibitem{datasetcollector}
J.~Leskovec, L.~A. Adamic, and B.~A. Huberman.
\newblock The dynamics of viral marketing.
\newblock volume~1, New York, NY, USA, May 2007. ACM.

\bibitem{EigenCollusion}
Q.~Lian, Z.~Zhang, M.~Yang, B.~Y. Zhao, Y.~Dai, and X.~Li.
\newblock An empirical study of collusion behavior in the maze p2p file-sharing
  system.
\newblock In {\em Proceedings of the 27th International Conference on
  Distributed Computing Systems}, ICDCS '07, pages 56--, Washington, DC, USA,
  2007. IEEE Computer Society.

\bibitem{CIKMInds}
E.-P. Lim, V.-A. Nguyen, N.~Jindal, B.~Liu, and H.~W. Lauw.
\newblock Detecting product review spammers using rating behaviors.
\newblock In {\em Proceedings of the 19th ACM international conference on
  Information and knowledge management}, CIKM '10, pages 939--948, New York,
  NY, USA, 2010. ACM.

\bibitem{mainWWW}
A.~Mukherjee, B.~Liu, and N.~Glance.
\newblock Spotting fake reviewer groups in consumer reviews.
\newblock In {\em Proceedings of the 21st international conference on World
  Wide Web}, WWW '12, pages 191--200, New York, NY, USA, 2012. ACM.

\bibitem{pagerank}
L.~Page, S.~Brin, R.~Motwani, and T.~Winograd.
\newblock The pagerank citation ranking: Bringing order to the web.
\newblock Technical Report 1999-66, Stanford InfoLab, November 1999.
\newblock Previous number = SIDL-WP-1999-0120.

\bibitem{cosine}
G.~Salton, C.~Buckley, and E.~A. Fox.
\newblock Automatic query formulations in information retrieval.
\newblock {\em Journal of the American Society for Information Science},
  34(4):262--280, 1983.

\bibitem{precision}
G.~Salton and M.~McGill.
\newblock {\em Introduction to modern information retrieval}.
\newblock McGraw-Hill computer science series. McGraw-Hill, 1983.

\bibitem{weight1}
Y.~Sun, Z.~Han, W.~Yu, and K.~Liu.
\newblock Attacks on trust evaluation in distributed networks.
\newblock In {\em Information Sciences and Systems, 2006 40th Annual Conference
  on}, pages 1461--1466. IEEE, 2006.

\bibitem{IEEE2102Survey}
Y.~Sun and Y.~Liu.
\newblock Security of online reputation systems: The evolution of attacks and
  defenses.
\newblock {\em Signal Processing Magazine, IEEE}, 29(2):87 --97, march 2012.

\bibitem{reliable}
G.~Swamynathan, K.~Almeroth, and B.~Zhao.
\newblock The design of a reliable reputation system.
\newblock {\em Electronic Commerce Research}, 10:239--270, 2010.
\newblock 10.1007/s10660-010-9064-y.

\bibitem{reptrap}
Y.~Yang, Q.~Feng, Y.~L. Sun, and Y.~Dai.
\newblock Reptrap: a novel attack on feedback-based reputation systems.
\newblock In {\em Proceedings of the 4th international conference on Security
  and privacy in communication netowrks}, SecureComm '08, pages 8:1--8:11, New
  York, NY, USA, 2008. ACM.

\bibitem{defending}
Y.~Yang, Y.~Sun, S.~Kay, and Q.~Yang.
\newblock Defending online reputation systems against collaborative unfair
  raters through signal modeling and trust.
\newblock In {\em Proceedings of the 2009 ACM symposium on Applied Computing},
  pages 1308--1315. ACM, 2009.

\bibitem{dishonest}
Y.-F. Yang, Q.-Y. Feng, Y.~Sun, and Y.-F. Dai.
\newblock Dishonest behaviors in online rating systems: cyber competition,
  attack models, and attack generator.
\newblock {\em J. Comput. Sci. Technol.}, 24(5):855--867, Sept. 2009.

\end{thebibliography}
\end{document}